\documentclass{elsart}

\usepackage{graphicx}

\usepackage{amssymb}

\begin{document}

\begin{frontmatter}



\title{Studies of the performance of different front-end systems for
flat-panel multi-anode PMTs with CsI(Tl) scintillator arrays}


\author[KYOTO]{H. Sekiya\corauthref{cor}},
\corauth[cor]{Corresponding author. tel:+81 75 753 3868; fax:+81 75 753 3799.}
\ead{sekiya@cr.scphys.kyoto-u.ac.jp}
\author[KYOTO]{K. Hattori},
\author[KYOTO]{H. Kubo},
\author[KYOTO]{K. Miuchi},
\author[WASEDA]{T. Nagayoshi},
\author[KYOTO]{H. Nishimura},
\author[KYOTO]{Y. Okada},
\author[KOBE]{R. Orito},
\author[KYOTO]{A. Takada},
\author[ICRR]{A. Takeda},
\author[KYOTO]{T. Tanimori},
\author[KYOTO]{K. Ueno}

\address[KYOTO]{Department of Physics, Graduate School of Science, Kyoto University, Kitashirakawa, Sakyo, Kyoto, 606-8502, Japan}

\address[WASEDA]{Advanced Research Institute for Science and
 Engineering, Waseda University, \\
17 Kikui-cho, Shinjuku, Tokyo, 162-0044, Japan}

\address[KOBE]{Department of Physics, Graduate School of Science and Technology, Kobe University, 1-1 Rokkoudai, Nada, Kobe, 657-8501, Japan}

\address[ICRR]{Kamioka Observatory, ICRR, University of Tokyo,\\
 456 Higasi-mozumi, Hida-shi, Gifu, 506-1205, Japan}

\begin{abstract}
We have studied the performance of two different types of front-end
 systems for our gamma camera based on 
Hamamatsu H8500 (flat-panel 64 channels multi-anode PSPMT)
with a CsI(Tl) scintillator array. The array consists of 64 pixels of
 $6\times6\times20{\rm mm}^3$ which corresponds to the anode pixels of H8500.

One of the system is based on commercial ASIC chips in order to
 readout every anode. The others are based on resistive charge divider network
between anodes to reduce readout channels.
In both systems, each pixel (6mm) was clearly resolved
by flood field irradiation of $^{137}$Cs.
We also investigated the energy resolution of these systems and
showed the performance of the cascade connection of resistive network
between some PMTs for large area detectors.
\end{abstract}

\end{frontmatter}

\section{Introduction}
\label{intro}
Recently, the concern with the gamma camera based on position sensitive PMTs for
application especially to nuclear medicine has been growing.   
The latest developed flat-panel type Hamamatsu H8500 and H9500\cite{HPK} are
promising devices for such purpose, 
and several studies have been conducted focusing on their spatial resolution 
with both pixellated scintillator array and continuous scintillator slab
aiming at PET and SPECT applications\cite{Pani,Gim,Herb}.

The merit of such multi-anode flat-panel type PMTs is the small non-active area 
when they are arrayed and constitute large area detectors, 
however, developments of readout systems for large number of 
channels are indispensable. 

On the other hand, Compton imaging detectors 
for gamma ray astronomy or next generation medical imaging
has been developed\cite{Takada,MEGA,medical} with gamma cameras used for the
detection of scattered gamma rays.
In such cases, not only the spatial resolution but also the 
energy resolution is important to reconstruct the direction of incident gamma rays.

In this paper,
we report the spatial resolution and energy resolution 
of our gamma camera based on H8500 with two different types of front-end
systems. One of the system is based on commercial ASIC chips in order to readout
every anode, the others are based on the resistive charge divider network
between anodes to reduce the readout channels.
In order to evaluate the performance, we coupled a CsI(Tl) scintillator array 
which fits to the anode pitches of H8500. 
This camera is intended for arrayed and covering our micro time
projection chamber (micro-TPC)\cite{Takada}, which constitutes a new Compton
imaging detector\cite{Tanimori}.

\section{The Detector}
The Hamamatsu H8500 has a very compact dimension of 52 mm $\times$ 52 mm $\times$ 28 mm
 with 12 stages of metal channel dynodes and a HV divider circuit.
The active photo cathode area is 49 mm$\times$49 mm and
 is covered by an 8$\times$8 anode array. 
The typical anode gain is $10^6$ (HV$=-1000$V) and the typical anode
 gain uniformity (the ratio of the maximum gain to the minimum gain) is about 2.5.
Each anode pixel size is 5.8 mm $\times$ 5.8 mm and the pitch between 
center of the anodes is 6.08 mm.

\begin{figure}[p]
\begin{center}
\includegraphics[width=5.5cm]{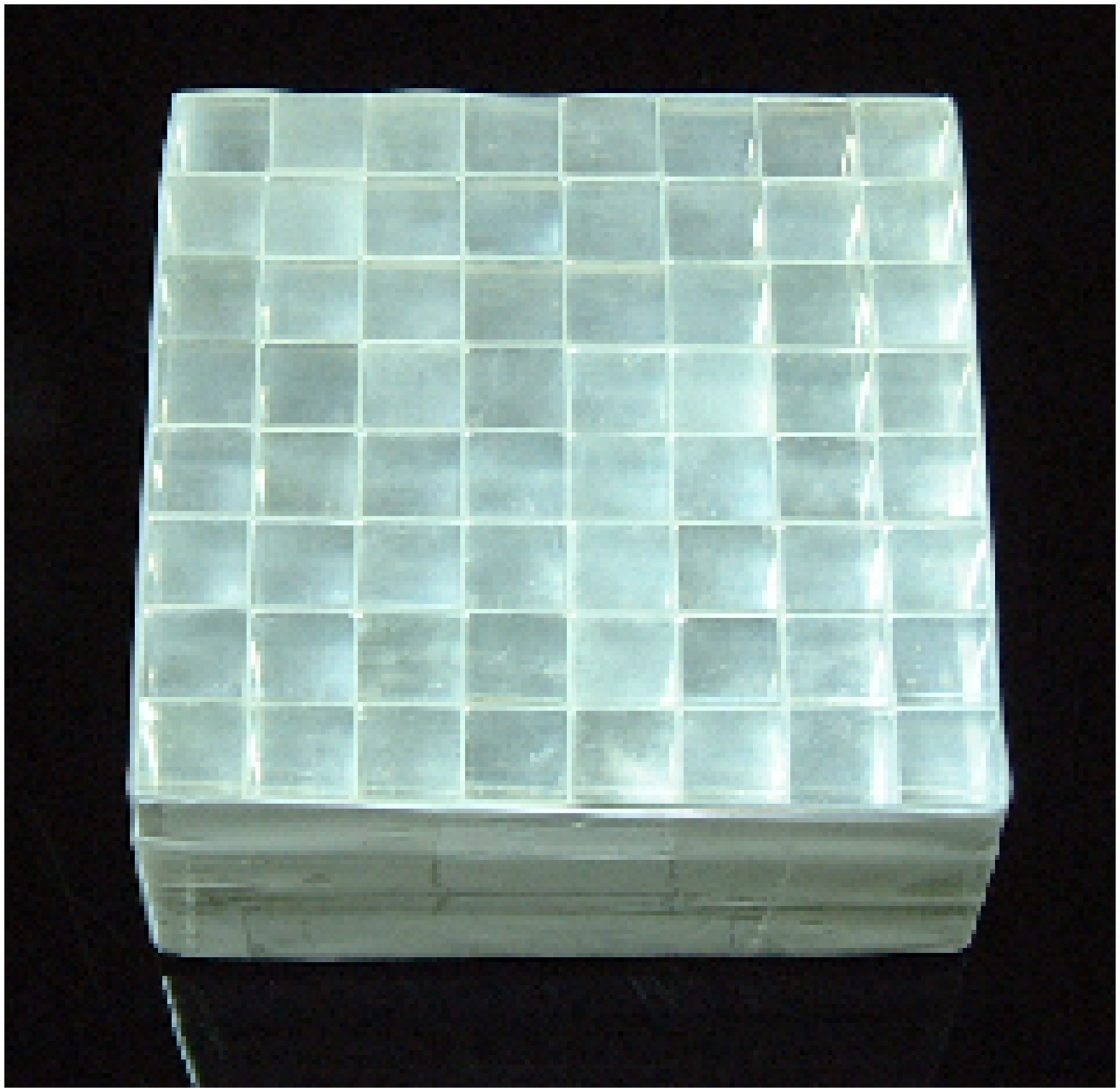}
\caption{Picture of the CsI(Tl) array.}
\label{fig:array}
\end{center}
\end{figure}

The size of each CsI(Tl) crystal is 6mm$\times$6mm$\times$20mm.
The crystals were also manufactured by Hamamatsu.
Between the crystals,
Vikuiti$^{\mbox{\scriptsize{\textcircled{\tiny R}}}}$ ESR
films (3M) of 65$\mu$m are inserted for the optical isolation,
so that the pixel of scintillator array corresponds to the anode pixel. 
The array is glued to H8500 using OKEN6262A optical grease.
Fig.\ref{fig:array} shows the picture of the array.

\section{Readout circuits}
\subsection{CP80068 system} 
\begin{figure}[p]
\begin{center}
\includegraphics[width=6.5cm]{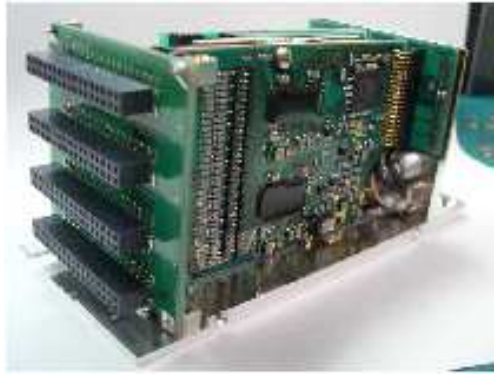}
\caption{Picture of CP80068.}
\label{fig:CP80068}
\end{center}
\end{figure}

Fig.\ref{fig:CP80068} shows
the individual anode readout system (Clear Pulse Co., Ltd. CP80068).
The dimension of CP80068 which is designed for 2 dimensional array of
H8500 is 52 mm $\times$ 52 mm $\times$ 95 mm.
It is based on two types of analog ASICs,
 VA32HDR14 and TA32CG2 manufactured by IDEAS ASA.
VA32HDR14 contains pre-amplifiers (input dynamic range$\sim\pm15$pC), 
shapers (gain$=118$mV/pC, peaking time$=2\mu$s), 
sample and hold circuits and a multiplexer.
TA32CG32 contains fast shapers (peaking time$=75$ns) and comparators, 
which can make the trigger signals.
The multiplexed 64ch data are digitized by a flash ADC on the CP80068
and sent to the VME sequence module via FPGAs.
It takes 164$\mu$s to process one event (64 channels).

\subsection{Resistive charge division}
\begin{figure}[p]
\begin{center}
\includegraphics[width=5.5cm]{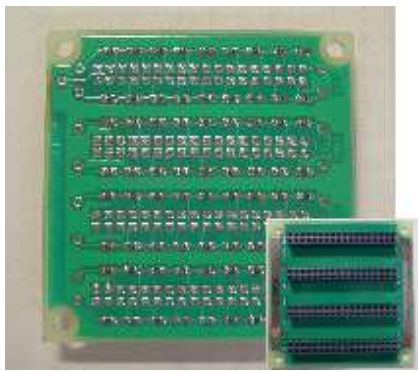}
\caption{Bottom view of the resistive divider network for H8500 and the
 top view(inset).}
\label{fig:RNB}
\end{center}
\end{figure}

Fig.\ref{fig:RNB} shows the charge divider network board for H8500.
Using this connector board, the anodes in horizontal rows of H8500 are connected
with 100$\Omega$ chips and the number of readout channels are reduced to 16.
Each reduced channel is preamplified (integrating time
constant$=66\mu$s), shaped (Clear Pulse CP4026, shaping time$=2\mu$s) and
digitized (CAEN V785). The last dynode output is used as the trigger signal. 

For further reduction of the readout channels,
 we connected the intervals of the the both edges of the
 horizontal chains with 100$\Omega$
resisters, thus 4 channels readout with resistive chain is also tested.

\section{Measurements and Results}
We are interested in 
the energy of sub-MeV region\cite{Takada}, accordingly,
the CsI(Tl) array was irradiated by 1 MBq $^{137}$Cs source (662keV)at a
distance of 30 cm. For the energy calibration, 
$^{22}$Na (511keV), $^{133}$Ba (356keV) and  $^{57}$Co (122keV) were also used. 
An important point to mention here is the dynamic ranges of the readout circuits.
As the input dynamic range of CP80068 is as small as $-15$pC,
H8500 should be operated with the gain of $10^4$ (HV$\sim600$V) 
to observe 662keV gamma rays. 
In the case of resistive charge division circuits,
dynamic ranges of the shaper and the ADC also 
limit the operation gain of H8500 to $10^5$ (HV$\sim800$V).

\subsection{Spatial Resolution}
\begin{figure}[p]
\begin{center}
\includegraphics[width=12.5cm]{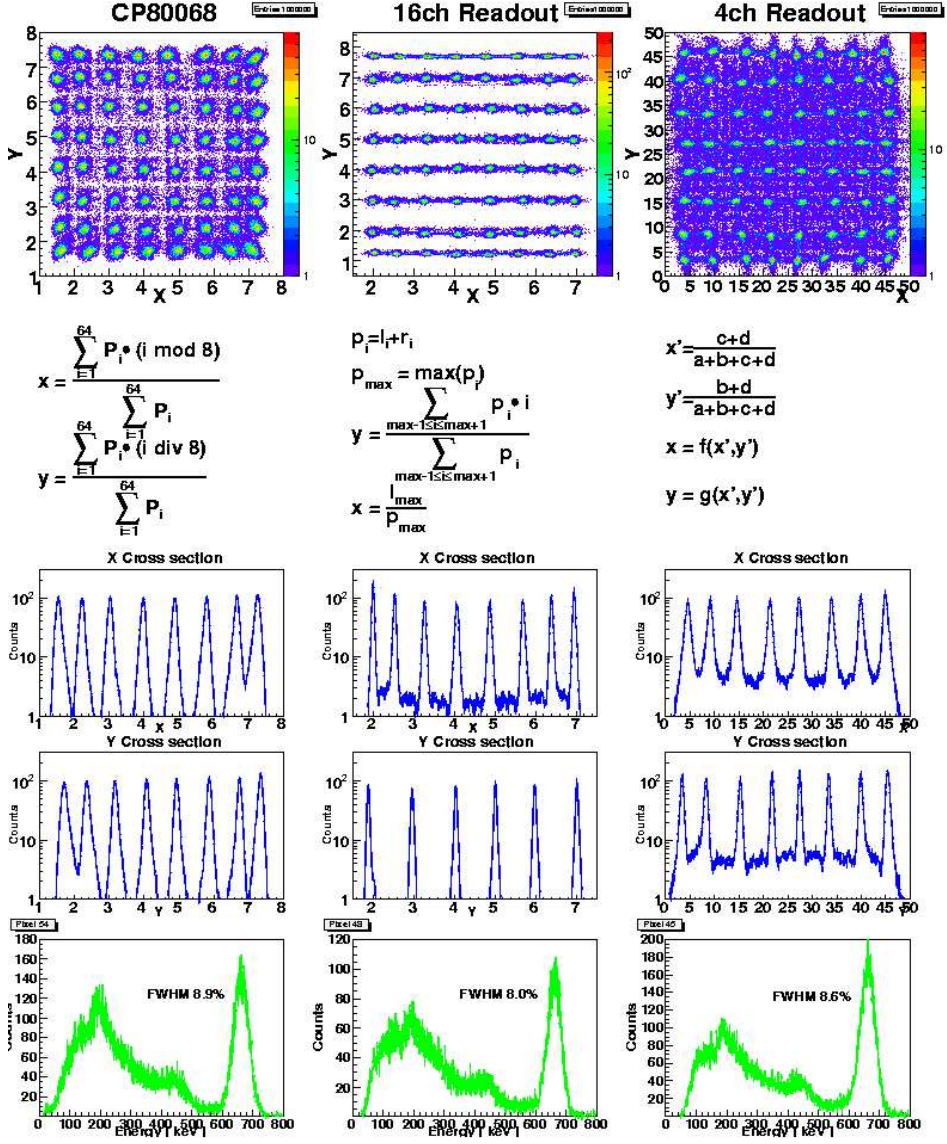}
\caption{Measurement results of each readout system. Flood field images
 of $^{137}$Cs irradiation, methods of the position reconstruction, x and y
 cross sections of central rows, the energy spectra of the best pixel of
 every readout system are shown. In the equations, $P_i$ is the ADC output of $i$th anode of
 CP80068 system, $l_i$($r_i$) is the ADC output of left(right) side of $i$th horizontal
 resistive chain of 16 channels readout system,
$a\cdot b\cdot c\cdot d$ represent the ADC outputs of 4 terminals of
 4 channels readout system.
In 4 channels readout system, as the raw image $(x',y')$ is distorted,
the corrected image $(x,y)$ calculated by TMultiDimFit class of
 ROOT\cite{ROOT} is indicated. }
\label{fig:res}
\end{center}
\end{figure}

The obtained flood irradiation images of $^{137}$Cs are shown in Fig.\ref{fig:res}.
The methods of the calculation of the position reconstruction are 
indicated as well. Image spots represent pixels of the CsI(Tl) array,
which indicates that the intrinsic spatial resolution of H8500 is
better than the anode pixel size.

The accidental hit events of multi pixels were rejected in the results of
CP80068 system (selection efficiency was $79\%$) and the accidental
hit events of more than two horizontal rows were also rejected in the 
results of the 16 channels readout system (selection efficiency was $85\%$).
On the other hand, in the 4 channels readout system, there is no way to
reject such events, therefore the peak to valley ratios of the x/y cross section
of the flood irradiation image are the worst. 

\subsection{Energy Resolution}
\begin{table}
\begin{center}
\begin{tabular}{c|c|c|c}  \hline
System         &  Best       & Typical       &   Worst     \\ \hline
CP80068        &   8.9\%     &  9.5\%        &  10.0\%     \\
16ch readout &   8.0\%     &  8.7\%        &   9.5\%     \\
4ch readout &   8.6\%     &  8.8\%        &   9.9\%     \\ \hline
\end{tabular}
\end{center}
\caption{Measured 662 keV energy resolutions (FWHM) of the pixels in each readout system.}
\label{tab:res}
\end{table}
The obtained energy spectra of the best pixel of each readout system are 
also shown in Fig.\ref{fig:res}.
The variations of the energy resolution of 662keV of
every readout system are summarized in Table.\ref{tab:res}.

\begin{figure}[p]
\begin{center}
\includegraphics[width=8cm]{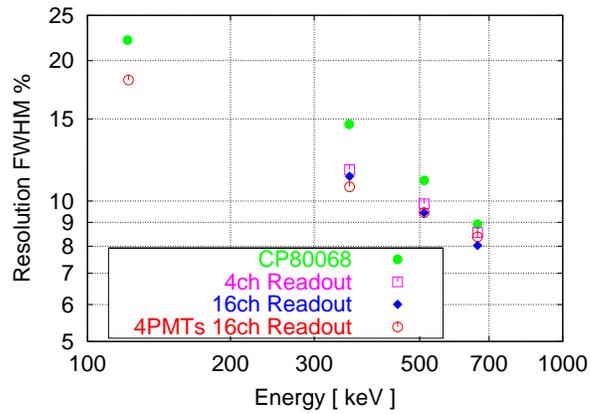}
\caption{Measured energy resolution of the best pixels of all the
 readout systems. Results of ``4PMTs 16ch Readout'' explained in Fig.\ref{fig:PMT4}
 are also indicated.}
\label{fig:res-ene}
\end{center}
\end{figure}

The variation of the resolution is mainly due to the variation of the
anode gain. Near the boundary of the detection area, optical leakage (photon collection
inefficiency) also affects not only the energy resolution but also the spatial
resolution.
Fig.\ref{fig:res-ene} shows the energy resolutions of measured energy of all the readout systems.
The reason why the energy resolution of the result of CP80068 system is the worst
is its low HV operation.

\section{Discussion and Conclusion}
It is admitted that individual anode readout is the best way for multi
anode PMTs, however, that needs development of 
exclusive ASICs with consideration for the 
light outputs of scintillator, gain of the PMT, and the dynamic range. 
Moreover, in our case, the spatial resolution is not determined by the anode pixel size
but by the crystal pixel size.

\begin{figure*}[p]
\begin{center}
\includegraphics[width=12.5cm]{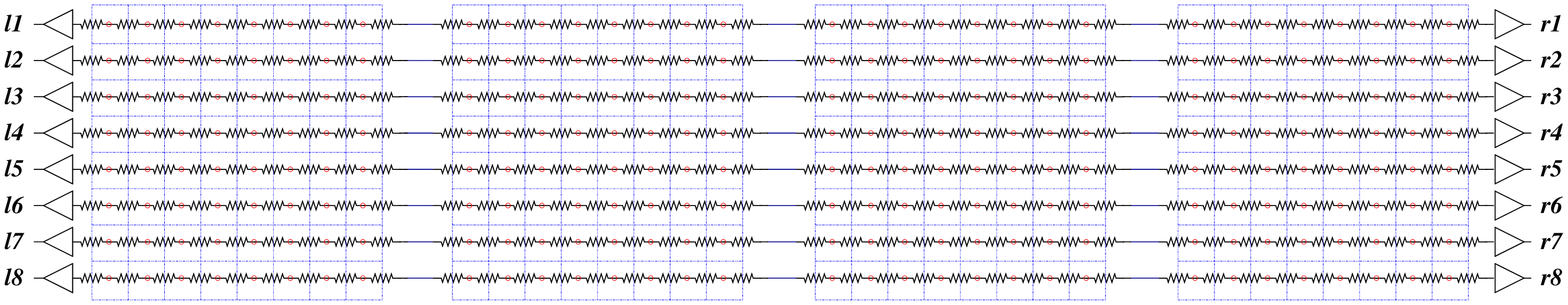}\\
\vspace{0.5cm}
\includegraphics[width=14.5cm]{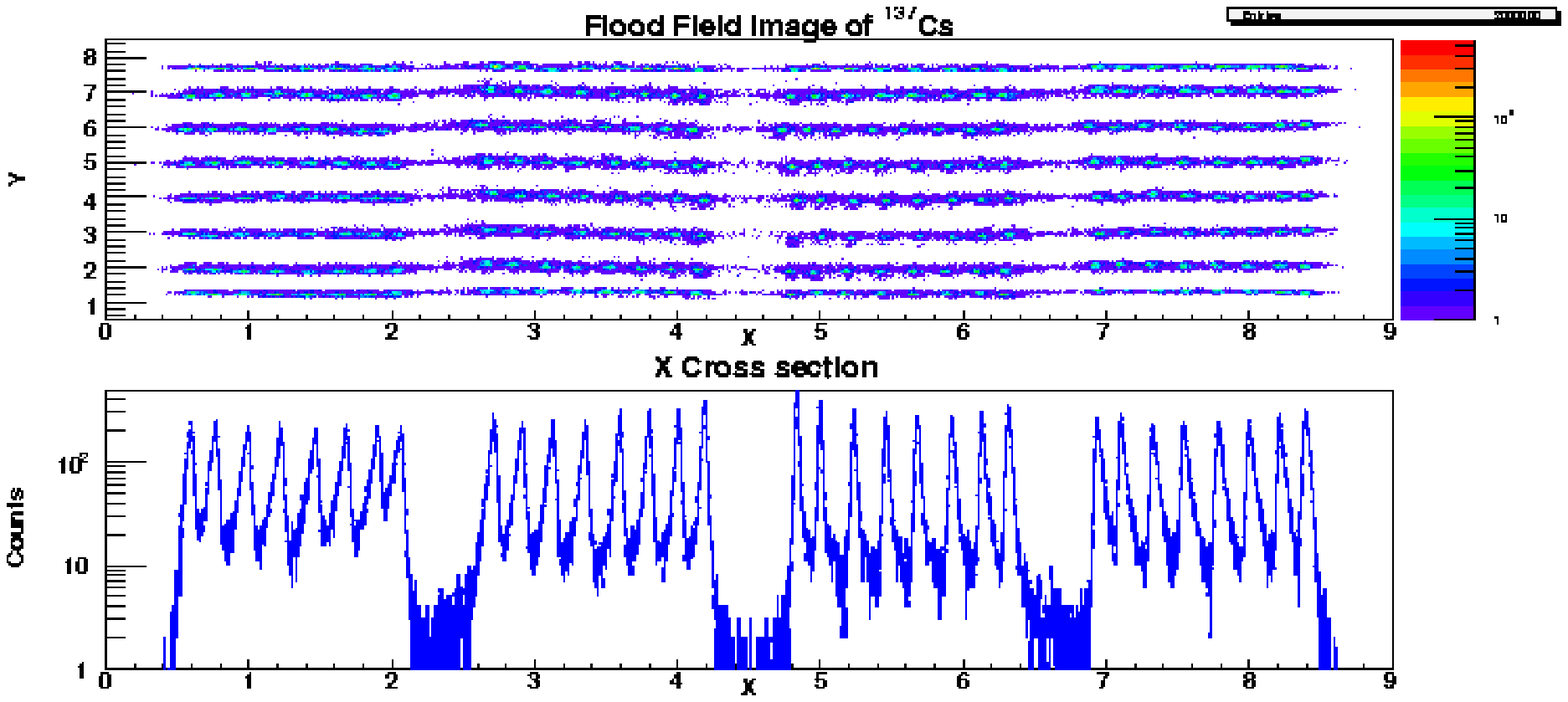}
\caption{Cascade connection of 4 H8500s with resistive charge divider network.}
\label{fig:PMT4}
\end{center}
\end{figure*}

Therefore, the advantage of energy resolution of the resistive charge 
divider network and discrete modules of readout circuit is encouraging 
to make larger area detector.
We made cascade resistive connection of 4 H8500s as shown in Fig.\ref{fig:PMT4} for
example. The energy resolution is also shown in Fig.\ref{fig:res-ene}
This connection is another example of 4 channels/PMT readout and
crystal pixel identification is better than that of previous 4 channels
readout system.   

In conclusion, large area detector of pixel scintillator and H8500 array with resistive charge
division systems have a good performance both energy and spatial
resolutions and have many possibilities in medical and gamma ray astronomy applications.

\section*{Acknowledgement}
\label{ack}
We would like to thank Takahashi Lab. at Institute of Space and 
Astronautical Science, Japan Aerospace Exploration Agency, Makishima
Lab. at Department of Physics, School of Scienece, University of Tokyo
and Dr. Gunji for the development of CP80068.



\end{document}